\documentclass[aps,prl,twocolumn,floatfix,showpacs]{revtex4}
\usepackage[dvips]{graphicx}
\usepackage{amssymb,amsfonts,amsmath}

\newcommand{\om}{\omega}
\newcommand{\Om}{\Omega}
\newcommand{\bv}{{\mathbf v}}
\newcommand{\bV}{{\mathbf V}}
\newcommand{\br}{{\mathbf r}}
\newcommand{\hatn}{\hat{n}}
\newcommand{\hatt}{\hat{t}}
\newcommand{\gat}{\gamma_a}
\newcommand{\dat}{\Delta}

\newcommand{\Oi}{\om_0}

\newcommand{\ri}{R_I}
\newcommand{\eps}{\epsilon}

\begin{document}

\title{Brownian ratchet in a thermal bath driven by Coulomb friction}

\author{Andrea Gnoli}
\affiliation{Istituto dei Sistemi Complessi - CNR, via del Fosso del Cavaliere
100, 00133 Rome, Italy}
\affiliation{Istituto dei Sistemi Complessi - CNR and Dipartimento di Fisica,
Universit\`a ''Sapienza'', p.le A. Moro 2, 00185 Rome, Italy}

\author{Alberto Petri}
\affiliation{Istituto dei Sistemi Complessi - CNR, via del Fosso del Cavaliere
100, 00133 Rome, Italy}

\author{Fergal Dalton}
\affiliation{Istituto dei Sistemi Complessi - CNR, via del Fosso del Cavaliere
100, 00133 Rome, Italy}

\author{Giacomo Gradenigo}
\affiliation{Istituto dei Sistemi Complessi - CNR and Dipartimento di Fisica,
Universit\`a ''Sapienza'', p.le A. Moro 2, 00185 Rome, Italy} 

\author{Giorgio Pontuale}
\affiliation{Istituto dei Sistemi Complessi - CNR, via del Fosso del Cavaliere
100, 00133 Rome, Italy}

\author{Alessandro Sarracino}
\affiliation{Istituto dei Sistemi Complessi - CNR and Dipartimento di Fisica,
Universit\`a ''Sapienza'', p.le A. Moro 2, 00185 Rome, Italy}

\author{Andrea Puglisi}
\affiliation{Istituto dei Sistemi Complessi - CNR and Dipartimento di Fisica,
Universit\`a ''Sapienza'', p.le A. Moro 2, 00185 Rome, Italy}

\begin{abstract}
 The rectification of unbiased fluctuations, also known as the ratchet
 effect, is normally obtained under statistical non-equilibrium conditions.
 Here we propose a new ratchet mechanism where a thermal bath solicits the
 random rotation of an asymmetric wheel, which is also subject to
 Coulomb friction due to solid-on-solid contacts. Numerical
 simulations and analytical calculations demonstrate a net drift
 induced by friction. If the thermal bath is replaced by a granular
 gas, the well known granular ratchet effect also intervenes, becoming
 dominant at high collision rates. For our chosen wheel shape the
 granular effect acts in the opposite direction with respect to the
 friction-induced torque, resulting in the inversion of the ratchet
 direction as the collision rate increases. We have realized a new
 granular ratchet experiment where both these ratchet effects are
 observed, as well as the predicted inversion at their crossover. Our
 discovery paves the way to the realization of micro and
 sub-micrometer Brownian motors in an equilibrium fluid,
 based purely upon nano-friction.
\end{abstract}


\maketitle

From microscopic organisms to muscle fibres, from electric motors to
power stations, the biosphere, our society and our lives critically
depend on the conversion of energy to mechanical work. Thermodynamics
provides precise and well established rules for energy conversion in
macroscopic systems but these rules become blurred at small scales
when thermal fluctuations play a decisive role~\cite{ref1}. Extracting
work under such conditions requires subtle strategies radically
different from those effective in the macroscopic
world~\cite{ref2,ref3,ref4,ref5,ref6,ref6a}. Within this framework, the
theory of Brownian motors deals with the rectification of thermal
fluctuations, a goal which can only be achieved in the presence of
dissipation~\cite{ref7,ref8,ref9,ref10,ref11,ref12}.  An interesting
class of systems, where both dissipation and fluctuations are
relevant, is represented by granular media~\cite{ref13,ref14}. Indeed,
interactions in a granular system are inherently dissipative, and
because of its small number of constituents when compared with
molecular gases or liquids, a granular fluid presents large
fluctuations. The additional break of spatial
symmetry is sufficient for a motor effect to be generated as
demonstrated in a series of experiments~\cite{ref15,ref16,ref17} and
theoretical works~\cite{ref18,ref19,ref20,ref21,ref22,ref23}.

In previous work the main source of dissipation was provided by the
inelasticity of collisions, a property normally not present at micro
or nanometric scales. The remarkable result of our study is a new
minimal model for a motor where energy is extracted from an
equilibrium bath and dissipated only through Coulomb
friction~\cite{persson}.  Friction is therefore demonstrated to be 
an unexpectedly efficient source of dissipation, that is able to rectify
unbiased fluctuations also in the case of {\em fully elastic}
collisions.  Such a model can therefore be exploited in micro and
  nano apparatuses where friction is still present~\cite{ref25}.

\begin{figure}[hp]
\includegraphics[clip=true,width=0.8\columnwidth]{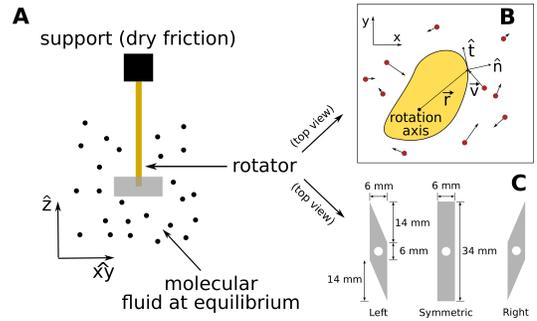}
\caption{A) Sketch of the model, front view. B) Top view, with
  explanation of quantities used in the text for a generic-shaped
  rotator. C) Top view, with specific shapes used in the simulations
  and in the experiment. }
\end{figure}

Our model, described pictorially in Fig. 1a, consists of: a 
wheel of mass $M$ and inertia $I$, rotating with angular velocity $\omega$ around a fixed axis (say $\hat{z}$).
The wheel is immersed in an equilibrium fluid and
collides with the molecules of mass $m$, and is
subject to a viscous drag $-\Gamma_{visc} \om$ and, most importantly, to
a Coulomb friction torque $-F_{friction} \sigma(\om)$ (where
$\sigma(x)$ is the sign function), due to solid-on-solid contacts within
its support, e.g. a spherical bearing. The equation of motion for the
angular velocity $\omega(t)$ of the wheel, therefore, reads
\begin{equation}
\dot{\om}(t)=-\gat \om(t) -\sigma[\om(t)]\dat + \eta_{coll}(t)
\end{equation}
where $\gat=\Gamma_{visc}/I$, $\dat=F_{friction}/I$ and
$\eta_{coll}(t)$ is the random force due to collisions with the
molecules of the bath.  The wheel is a cylinder parallel to the
rotation axis $\hat{z}$. Its base in the plane $\hat{xy}$ (shown in
Fig. 1b for a generic shape) can be symmetric or asymmetric for
inversion of one of the two axis ($\hat{x}$ or $\hat{y}$). The two
specific shapes taken in consideration here, one symmetric and other
asymmetric, are drawn in Fig. 1c.  The velocities of the molecules are
distributed according to the Maxwell-Boltzmann distribution, with the
only parameter being the ``thermal'' velocity $v_0=\sqrt{\langle v^2
  \rangle}$, where $v$ is a component of the velocity vector on the
$\hat{xy}$ plane. The molecular bath is also characterized by its
number density $n$. The main parameter for the collision between the
wheel and the molecules is the total cross section $\Sigma$. The
collision rule, given in details in the Supplemental
Material~\cite{sm}, conserves total angular momentum and may dissipate
part of the total kinetic energy, according to the value of the
restitution coefficient $\alpha \in [0,1]$. We will show that our
model exhibits the ratchet effect even in the case of fully elastic
collisions ($\alpha=1$), and even if the viscous force is removed
($\gamma_a=0$). The choice of a more general (possibly inelastic)
collision rule and the presence of a weak viscous damping is necessary
to account for the results of the granular experiment described
below. For consistency with the experiment, the viscous force (if
present) is assumed to be small enough that $\gamma_a |\omega| \ll
\Delta$ for most of the values of $\omega$.

When the range of interactions with the molecules is short enough (as in
the hard-core case), only two time-scales are relevant
in the system: 1) the mean stopping time due to environmental
dissipation, dominated by Coulomb friction, $\tau_\Delta=\frac{\langle
  |\omega|\rangle_{pc}}{\dat} \sim \frac{\eps v_0}{R_I \dat}$, where
$\langle \cdot \rangle_{pc}$ denotes a post-collisional average,
$\ri=\sqrt{I/M}$ is the radius of inertia and $\eps=\sqrt{m/M}$; 2)
the mean free time between two collisions $\tau_c \sim \frac{1}{n
  \Sigma v_0}$. We therefore use as main control parameter
\begin{equation}
\beta^{-1}=\frac{\eps n\Sigma v_0^2}{\sqrt{2} \pi \ri \dat}  \approx
\frac{\tau_\Delta}{\tau_c}
\end{equation}
which is an estimate of the ratio of those two time-scales, as
verified by simulations~\cite{sm}.  As
already noticed~~\cite{ref21,ref22}, when $\beta^{-1} \gg 1$ ($\tau_c
\ll \tau_\Delta$) the dynamics of the rotator is dominated by
collisions and friction is negligible (frequent collisions limit,
denoted in the following by FCL); in the opposite limit $\beta^{-1}
\ll 1$ ($\tau_c \gg \tau_\Delta$, rare collisions limit, RCL) the
rotator remains most of the time at rest and is rarely perturbed by
collisions acting as independent random excitations.

The complex behaviour of the model is simplified in
the diluted limit, when Molecular Chaos can be assumed. With such an
assumption, the probability density function (pdf) $p(\omega,t)$ for
the angular velocity is fully described by the following linear
Boltzmann equation~\cite{vk,ref23,ref21}
\begin{subequations} \label{beq}
\begin{align} 
\partial_t p(\omega,t)&=\partial_\omega [(\dat \sigma(\omega)+\gat
\omega)p(\omega,t)]+J[p,\phi] \\
J[p,\phi]&=\int d\omega' W(\omega|\omega')p(\omega',t)- p(\omega,t) f_c(\omega),
\end{align}
\end{subequations}
where we introduce the rate $W(\omega'|\omega)$ for the transition
$\omega \to \omega'$ and the velocity-dependent collision frequency
$f_c(\omega)=\int d\omega' W(\omega'|\omega)$. The rate
$W(\omega'|\omega)$, given explictly for hard-core interactions in~\cite{sm}, depends on the velocity distribution of the gas
particles, on the restitution coefficient $\alpha$, on the rotator
cross section and on the density of the gas.

\begin{figure}
\includegraphics[clip=true,width=0.8\columnwidth]{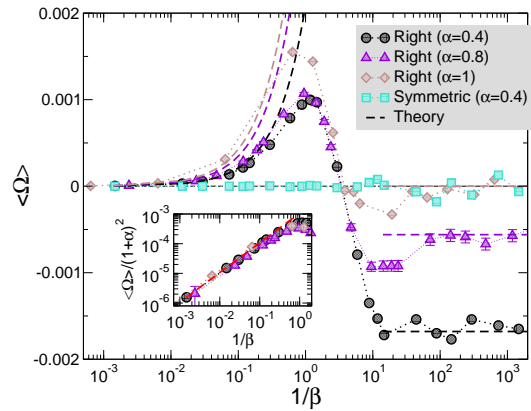}
\caption{Simulations under the assumption of Molecular Chaos. The
  rescaled average angular velocity $\langle \Omega \rangle$ is shown
  as a function of $\beta^{-1}$. Theoretical expectations in the FCL
  and RCL are marked by dashed lines. The lower inset zooms in the RCL
  region. Simulations are performed using
  shapes and dimensions of Fig. 1c, $\Delta=10$, $\gamma_a=0$,
  $R_I/\epsilon=10^3$ and $v_0=100$ in arbitrary units (varying $n
  \Sigma$ to obtain a variation of $\beta^{-1}$).  }
\end{figure}

A first insight into Eq.~\eqref{beq} is obtained by Direct Simulation
Monte Carlo~\cite{ref31,sm}, whose results are summarized in Fig.~2,
always keeping $\gamma_a=0$. The Figure shows the average velocity of
the ratchet rescaled by the ideal ``thermal'' velocity,
i.e. $\langle\Omega\rangle = \frac{R_I}{\epsilon
  v_0}\langle\omega\rangle$, for several values of $\alpha$ and
different shapes. Our main, unprecedented, result is the existence of
an average drift, i.e. a motor effect, in the case of elastic
collisions, provided that the shape is asymmetric (curve with diamond
symbols). This effect is independent of the presence of the viscous
damping, which very weakly affects the results of the simulation. In
the elastic case, the average drift disappears for large $\beta^{-1}$,
where the effect of friction is washed out by highly frequent
collisions and the system equilibrates with the bath.  Remarkably, the
ratchet effect starting from the RCL increases, in absolute value,
when $\beta^{-1}$ increases, so that it must go through a maximum. We
interpret such a maximum as a kind of {\em stochastic resonance}: the
rotator switches from the ``drift'' state to the ``rest'' state on the
time-scale $\tau_\Delta$ and switches back to the ``drift'' state on
the time-scale $\tau_c$. When the two scales synchronize with each
other, the total time spent in the ``drift'' state is maximized, as
well as its average velocity.

As demonstrated in previous studies~\cite{ref18,ref19}, in the inelastic
case ($\alpha<1$) a ratchet effect survives also in the FCL:
interestingly, it takes different signs with respect to the 
RCL. Therefore, the crossover between these two limits requires
the presence of an inversion point. 

It is possible to have an analytical account of the two opposite RCL and FCL limits~\cite{ref23,ref21}. 
When the mass of the rotator is large, $\eps \ll 1$  the FCL is perturbatively reduced to a Brownian
approximation and the average drift has already been computed~\cite{ref23} giving
\begin{subequations} \label{fcl}
\begin{align}
\langle \Om \rangle = \eps
\sqrt\frac{\pi}{2}\frac{1-\alpha}{4} \mathcal{A}_{FCL}\\
\mathcal{A}_{FCL}=-\frac{\langle g^3 \rangle_{surf}}{\langle g^2
  \rangle_{surf}},
\end{align}
\end{subequations}
where the asymmetry of the rotator is represented by
$\mathcal{A}_{FCL}$ which is $0$ for a symmetric rotator;
above we have used the shorthand notation for the uniform average
along the perimeter (denoted as ``surface'') of the base of the wheel
 $\langle  \rangle_{surf} = \int_{surf} 
\frac{ds}{S}$ ($S$ being the total perimeter~\cite{sm}), while
$g=\frac{{\mathbf r}\cdot \hat{t}}{\ri}$; see Fig.~1b for an
explanation of symbols. This formula predicts zero net drift either
with elastic collisions ($\alpha=1$) or with a symmetric rotator
($\mathcal{A}_{FCL}=0$ ), as expected from symmetry arguments. Most
importantly, it predicts a constant value, as
verified in the numerical simulations. This implies $|\langle \omega \rangle| \sim
v_0$ for the {\em dimensional} angular velocity.

The study of the RCL leads to remarkably different predictions.  In
such a limit, the dynamics after each collision event produces an
increment of the angular position of the rotator $\Delta \theta$ which
depends on the velocity $\bv$ of the gas particle, precisely on its
projection $v=\bv \cdot \hatn$, and on the point of impact represented
by its curvilinear abscissa $s$. The formula is $\Delta
\theta(v,s)=\sigma(\Oi)\frac{\Oi^2}{2\dat}$ with
$\Oi=-(1+\alpha)\frac{v}{\ri}\frac{\eps^2 g}{1+\eps^2 g^2}$.
Following the calculations detailed in~\cite{sm}, one obtains the formula for
the rescaled average velocity of the ratchet
\begin{subequations} \label{rcl}
\begin{align} \label{rcla}
\langle \Omega \rangle  &=  \sqrt{\pi}(1+\alpha)^2\beta^{-1} \eps^2 \mathcal{A}_{RCL}\\
\mathcal{A}_{RCL} &= \left\langle
\frac{ \sigma(g)g^2}{(1+\eps^2 g^2)^2} \right\rangle_{surf},
\end{align}
\end{subequations}
where $\mathcal{A}_{RCL}=0$ for symmetric shapes of the rotator.
Eq.~\eqref{rcl} shows that a non-zero drift is achieved {\em for any
  value of the restitution coefficient}: even in the (ideal) elastic
case, Coulomb friction alone produces the desired ratchet effect
provided that the shape of the rotator is not symmetric, i.e. that
$\mathcal{A}_{RCL} \neq 0$. Note that the limit of vanishing dry
friction ($\Delta \to 0$) is singular in formula~\eqref{rcla}, since
in the absence of dissipation between collisions the stopping time
becomes infinite, $\tau_\Delta \to \infty$, and the assumption of
``rare collisions'' breaks down.  Equally interesting, the shape
factor $\mathcal{A}_{RCL}$, determining the intensity and drift
direction in the RCL, can take {\em opposite sign} with respect to the
shape factor $\mathcal{A}_{FCL}$ in the FCL formula. This is precisely
the case for our chosen shape, see Fig.~1c. Moreover, the magnitude of
the drift is predicted to increase with $\beta^{-1}$ as seen in the
numerical simulations for small $\beta^{-1}$. This corresponds to
$|\langle \omega \rangle| \sim v_0^3$. Both the predictions for the
RCL and for the FCL are superimposed on the results (elastic and
inelastic) of the numerical simulations in Fig. 2, demonstrating
excellent agreement in their respective limits. We remark that if
friction is removed ($\Delta=0$) the only ratchet effect is the one predicted in Eq.~\eqref{fcl}, i.e. no inversion is observed.

\begin{figure}[t!]
\includegraphics[clip=true,width=0.3\columnwidth]{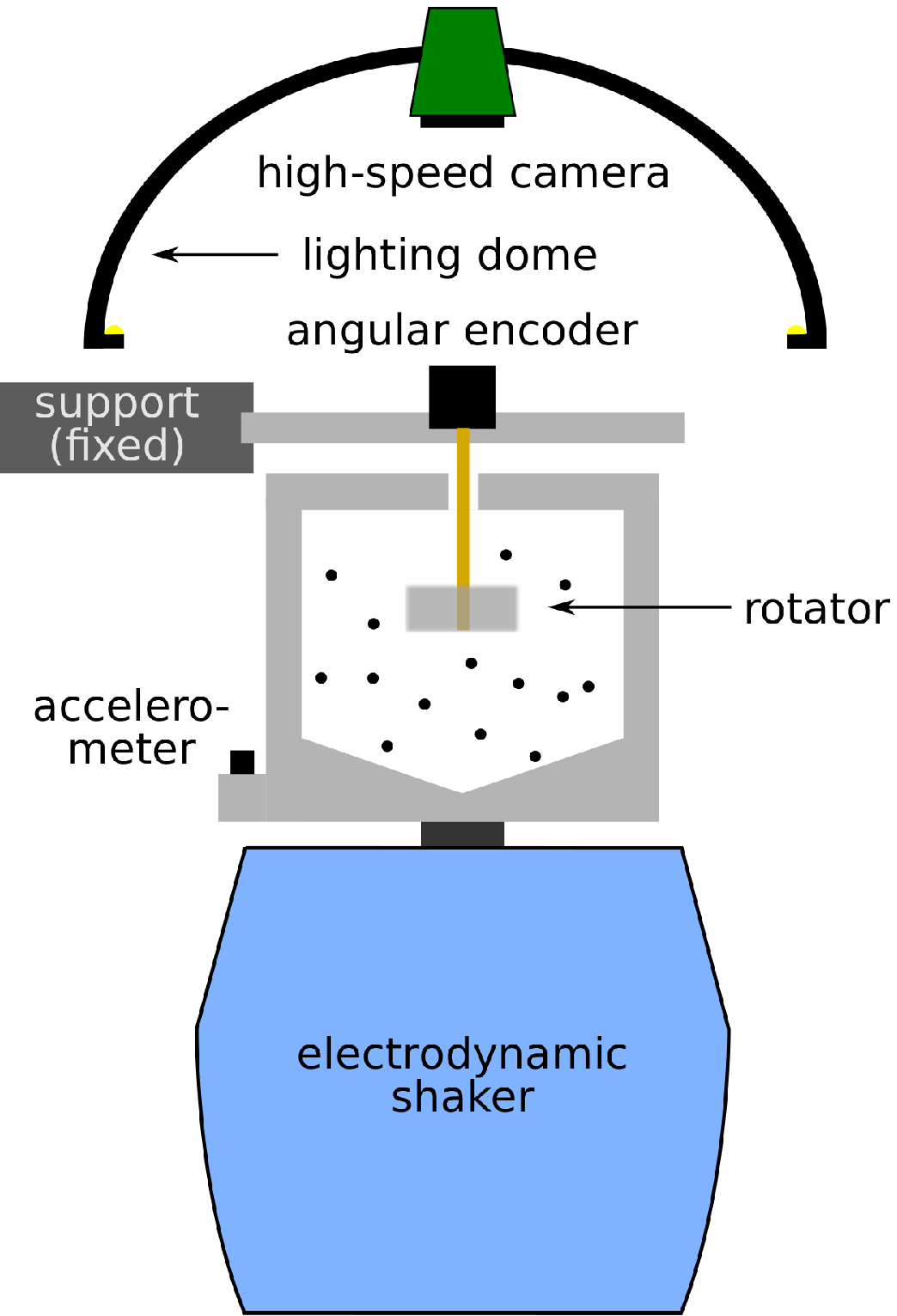}
\includegraphics[clip=true,width=0.65\columnwidth]{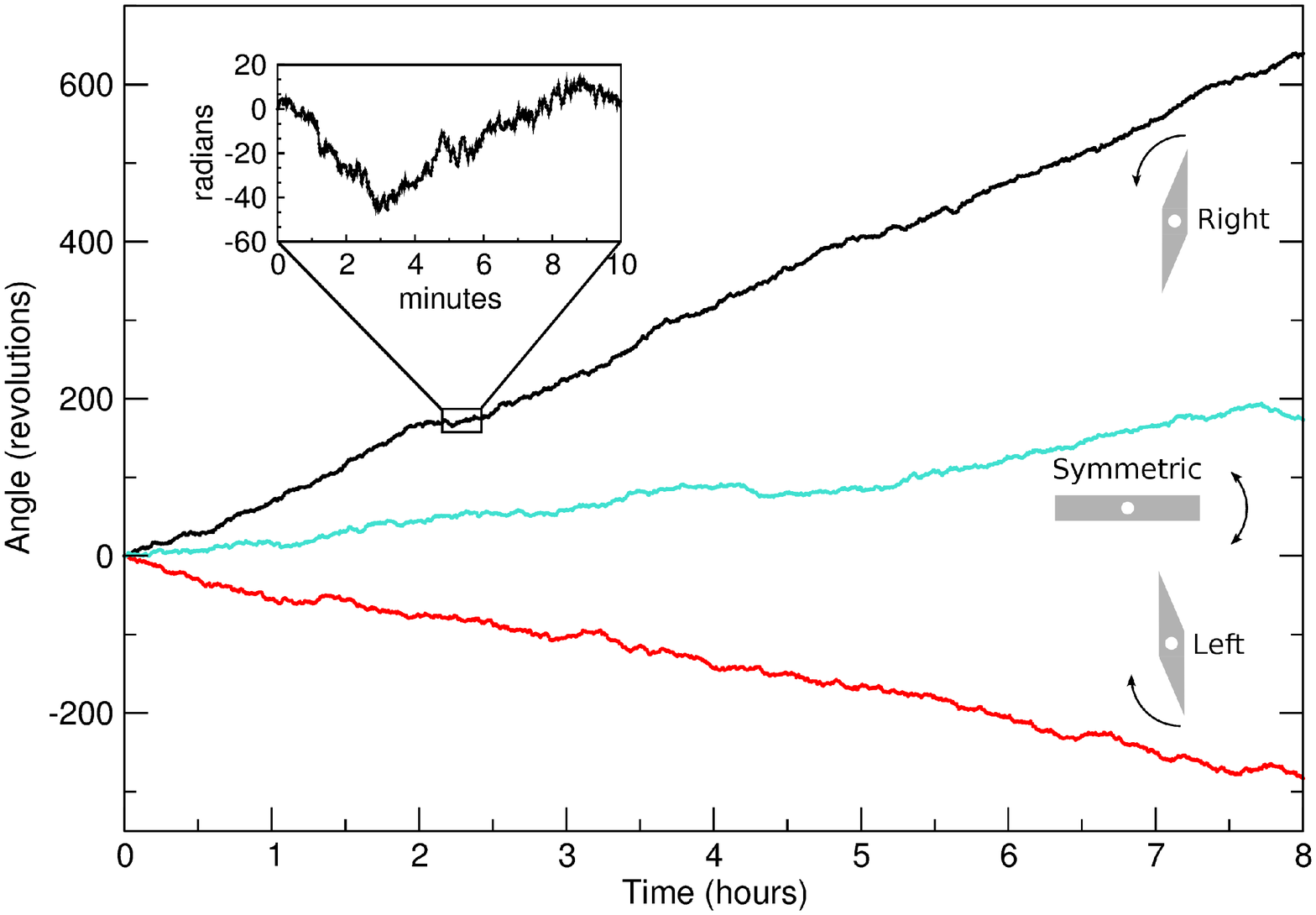}
\caption {Real experiment. A) Experimental setup. B) The angular position of the rotator as a
  function of time, highlighting the ratchet effect for the asymmetric
  rotator.  The inset shows an enlargement (ten minutes) extracted
  from a asymmetric (right) experiment. For these experiments the
  value of maximum acceleration, normalized by gravity, is $13$.}
\end{figure}

In order to obtain the first experimental evidence of this newly discovered
ratchet effect, we have built a macroscopic realization of our model, i.e. a
setup where the thermal bath is replaced by a fluidized granular
gas. In such a setup the collisions are unavoidably inelastic:
nonetheless, by tuning the collision frequency, it is possible
to disentangle the two ratchet mechanisms which act in opposite
directions, and so demonstrate the newly discovered effect induced by Coulomb friction. Our setup consists of a rotator
vertically suspended in a granular medium (see Fig.~3a) maintained
by an electrodynamic shaker in a (roughly homogeneous) stationary
gaseous regime~\cite{ref14,ref27}.  The statistics of the velocities
of the grains, on the rotation plane, has been verified to be
indistinguishable from a Maxwell-Boltzmann distribution~\cite{sm}.  The shaker
performs a vertical sinusoidal oscillation at a frequency of $53$ Hz,
while the amplitude is varied to explore different regimes of the
system.  We stress that the rotator is not in direct contact with the
shaker, it only collides with the flying grains. Its motion is
recorded by an angular encoder which also supports it
through two precision spherical bearings.  Two rotators have been
realized to reproduce the two different shapes of Fig. 1c. The
asymmetry of the latter can be inverted (from left to right hand, and
vice versa) simply by turning the rotator upside-down.  The amplitude
of the shaker oscillation is varied to span a range of maximum
acceleration (in units of gravity acceleration) between $5$ to $20$,
resulting in a granular thermal velocity $v_0$ ranging between $\sim
100$ mm/s and $\sim 600$ mm/s. Details of the experimental setup and
measurement of the parameters are in~\cite{sm}.

In Fig.~3b, we provide some runs demonstrating the ratchet effect for
the asymmetric rotator. A net drift is evident when a chiral rotator
is used. Turning the rotator upside down results in a reversed direction of
rotation.  The symmetric rotator makes a diffusive motion with only a
small drift revealing the presence of some bias due to imperfections
in the rotational symmetry of the setup. Such a bias affects also the
curves pertaining to the asymmetric rotator and it does not appear to
depend on the rotator's direction of asymmetry, or absence
thereof. The typical instantaneous velocity of rotators goes from $\sim
10^{-1}$ to $\sim 1$ rad/s, much larger than the typical ratchet drift, which goes from 
$\sim 10^{-4}$ to $10^{-2}$ rad/s. Such a low signal-to-noise ratio
makes it impossible to determine the true behaviour of the rotator if
one monitors its position only for short times (e.g. less than a few
hours), as evidenced in the inset of Fig.~3b.

\begin{figure}[t]
\includegraphics[clip=true,width=0.9\columnwidth]{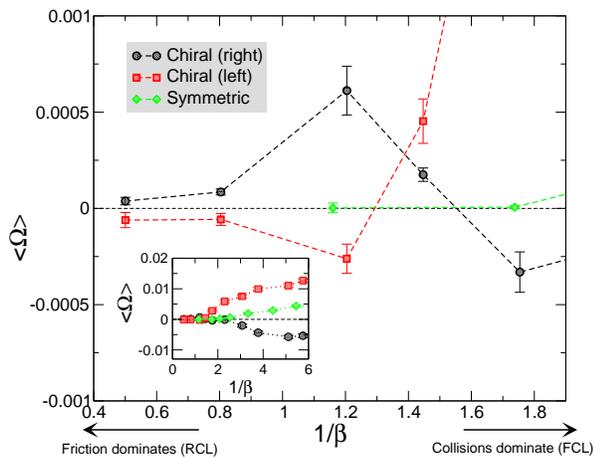}
\caption{Real experiment. Plot of the rescaled angular velocity of the rotator,
$\langle\Omega\rangle = \frac{R_I}{\epsilon v_0}\langle\omega\rangle$,
averaged on an $8$-hours run, as a function of $\beta^{-1}$ which estimates the ratio of time-scales
$\frac{\tau_\Delta}{\tau_c}$. The main plot shows the region where a
maximum and a current inversion are observed for the asymmetric
rotator. The inset sums all the results up, highlighting the behaviour
for large $\beta^{-1}$. Experiments are performed with maximum shaker acceleration
varying in the range from $5$ to $20$ in gravity units.}
\end{figure}

In Fig.~4, we show the average adimensional angular velocity
$\langle \Om \rangle=\frac{\ri}{\eps v_0}\langle\om\rangle$ for a set
of experiments with asymmetric and symmetric rotators. The behaviour
with $\beta^{-1}$ strongly resembles that observed in our numerical
simulations (Fig. 2). At $\beta^{-1}<1$, we measure $|\langle \Om
\rangle|$ increasing with $\beta^{-1}$, followed by a maximum in the
proximity of $\beta^{-1} \sim 1$ and then by an inversion of direction
of motion.  At large $\beta^{-1}$, $|\langle \Om \rangle|$ increases
and finally reaches a plateau (see the inset of Fig.~4). Even with
some quantitative discrepancies, we can claim that our experiment
reproduces very well the qualitative phenomenology of the model,
including the resonant maximum and the inversion point, which are both
evidence of the presence of the friction-induced ratchet
mechanism. We believe that the quantitative differences (the real
ratchet is faster roughly by a factor $2$ in the RCL and a factor $5$
in the FCL) can be imputed to the many assumptions present in the
model, the most important being molecular chaos and spatial
homogeneity, hardly controlled in the experiment: indeed
non-equilibrium correlations may well become important at high collision
frequencies~\cite{ref32}.

To conclude, our main discovery is the existence of a minimal
  ratchet model made of two simple ingredients: a wheel subject to
  Coulomb friction and a bath at thermodynamic equilibrium. Such a model
  appears even simpler than the classical Feynmann-Smoluchowsky
  model~\cite{ref8}.  The observation in the laboratory of a maximum and an
  inversion of the ratchet velocity (Fig.~4), due to the crossover
  from the inelasticity-dominated (FCL) to the friction-dominated
  (RCL) regime, is a strong experimental demonstration of the
  efficiency of this effect. We wish to remark that, in all previous experimental
  and theoretical work on friction-driven
  ratchets~\cite{eglin,buguin,ref26,ref24b}, the energy injection was
  provided by mechanisms different from an equilibrium bath: our
  proposal is the first which can be realized at the micro- and
  nano-scale in an equilibrium fluid, that is, without the application
  of any external field.

\begin{acknowledgments}
We would like to thank A. Vulpiani for useful comments and MD. Deen
for technical support. The authors acknowledge the support of the
Italian MIUR under the grants: FIRB-IDEAS n. RBID08Z9JE, 
FIRBs n. RBFR081IUK and n. RBFR08M3P4, and PRIN  n. 2009PYYZM5.
\end{acknowledgments}

\section{SUPPLEMENTAL MATERIAL}

\subsection{Details of the theory}

We recall here some of the main quantities of our model (refer to Figure 1 for a visual
explanation of symbols).  A rigid body of momentum of inertia $I$ and
total mass $M$ is bound to rotate around a fixed axis (say $\hat{z}$)
and is suspended in a molecular or granular fluid. The body is
constituted by the set of material points with cartesian coordinates
$\{x,y,z\}$ with $z\in[0,h]$ (where $h$ is the height of the cylinder)
and $\sqrt{x^2+y^2}<r(s)$ for each $s\in[0,S]$ where $s$ is the
curvilinear abscissa, $r(s)$ is the curve delimiting a section of the
solid in the $\hat{xy}$ plane, and $S$ is the perimeter of the
section.  The fluid surrounding the rotator has number density $n$ and
is made of identical spheres of mass $m$. We denote by $\omega$ the
angular velocity of the rotator, by $\theta$ its angular position, and
by $\bv$ the velocity of a molecule of the fluid. We also denote by $\rho=n
h$ the two-dimensional projection of density, which is the only one
which matters in our problem. Note that $\rho S \equiv n \Sigma$ if
$\Sigma$ (as in the Letter) is the total scattering cross section of the rotator
against the molecules of the bath.

{\em Inelastic collisions}
The fluid interact with the rotator by means of inelastic
collisions, which change the velocity of the rotator
from $\omega$ (before the interaction) to $\omega'$ (after the interaction) and that of the colliding particle from
$\bv$ to $\bv'$. In the case of hard-core collisions the scattering event follows the rules
\begin{subequations} \label{col_rule}
\begin{align} 
\omega'&=\omega+(1+\alpha)\frac{(\bV- \bv) \cdot \hatn}{\ri}\frac{g\eps^2
}{1+\eps^2 g^2},\\
\bv' &= \bv + (1+\alpha)\frac{(\bV- \bv) \cdot \hatn}{1+ \eps^2 g^2}\hatn
\end{align}
\end{subequations}
where $\alpha \in [0,1]$ is the restitution coefficient ($\alpha=1$
corresponds to elastic collisions), ${\mathbf V}=\omega \hat{z} \times
{\mathbf r}$ is the linear velocity of the rotator at the point of
impact ${\mathbf r}$, $\hat{n}$ is the unit vector perpendicular to
the surface at that point, and finally $g=\frac{{\mathbf r}\cdot
  \hat{t}}{\ri}$ with $\hat{t}=\hat{z} \times \hat{n}$ which is the
unit vector tangent to the surface at the point of impact.

Equations~\eqref{col_rule} guarantee that total angular momentum
$L \hat{z}=m \br \times \bv+I \omega \hat{z}$ is conserved, that relative
velocity projected on the collision unit vector is reflected and
rescaled by the restitution coefficient, $(\bV'- \bv') \cdot \hatn=-\alpha (\bV-
\bv) \cdot \hatn$,
and finally that the kinetic energy
$K=\frac{m}{2}|\bv|^2+\frac{I}{2}\omega^2$ changes as
\begin{equation}
K'-K=-\frac{(1-\alpha^2)}{2}[(\bV- \bv) \cdot \hatn]^2\frac{m}{1+\eps^2 g^2}.
\end{equation}
A few relations in cartesian coordinates may be useful: $\bV=(-\omega
r_y, \omega
r_x)$ and $\hatt=(-n_y,n_x)$. It is also useful to realize that $\bV \cdot
\hatn=-\omega \ri g$.

{\em Transition rates for a Gaussian gas}
The transition rate for the collisional Markov process reads~\cite{ref23}
\begin{subequations}
\begin{align}
W(\omega'|\omega)&=\rho S \int \frac{ds}{S} \int d\bv \phi(\bv) \Theta[(\bV-
\bv) \cdot \hatn] \times \\ &|(\bV- \bv) \cdot \hatn|\delta[\omega'-\omega-\Delta
\omega(s)],\\
\Delta \omega(s) &= (1+\alpha)\frac{[\bV(s)- \bv] \cdot \hatn}{\ri(s)}\frac{
g(s)\eps^2}{1+\eps^2g(s)^2},
\end{align}
\end{subequations}
where $\phi(\bv)$ is the pdf for the gas particle velocities and the
Heaviside step function $\Theta[(\bV- \bv) \cdot \hatn]$ enforces the
kinematic condition necessary for impact.

Assuming that the velocities of the molecules of the surrounding fluid obey the Maxwell-Boltzmann statistics
$\phi(\bv)=\frac{1}{2\pi
  v_0^2}e^{-\frac{v^2}{2 v_0^2}}$, one gets an explicit expression
for the transition rates 
\begin{multline}
W(\omega'|\omega)=\frac{\rho S \ri^2}{(1+\alpha)^3 \sqrt{2\pi} \eps^2 v_0}\int
\frac{ds}{S} |\omega'-\omega|\frac{(1+\eps^2g^2)^2}{\eps^2g^2} \\
\times\Theta\left[\frac{\omega'-\omega}{g}\right]\exp\left[-\frac{\ri^2}{2\eps^2
v_0^2} \left(\omega \eps g+\frac{(\omega'-\omega)(1+\eps^2g^2)}{(1+\alpha)\eps
g}\right)^2\right].
\end{multline}

{\em Average velocity in the RCL}
When $\beta^{-1} \ll 1$ ($\tau_\Delta \ll \tau_c$, denoted as Rare
Collisions Limit, RCL), the dynamics is dominated by the rotator at
zero velocity with random and independent perturbations due to sparse
collisions with gas particles~\cite{ref21}. In this case, the dynamics after each
collision event produces an increment of the angular position of the
rotator $\Delta \theta$ which depends on the velocity $\bv$ of the gas
particle, precisely on its projection $v=\bv \cdot \hatn$, and on the
point of impact represented for instance by its curvilinear abscissa
$s$. The formula in the limit of negligible viscous damping ($\gamma_a \to 0$) is
\begin{align}
\Delta \theta(v,s)&=\sigma(\Oi)\frac{\Oi^2}{2\dat}\\
\Oi&=-(1+\alpha)\frac{v}{\ri}\frac{\eps^2 g}{1+\eps^2 g^2}.
\end{align}
As a consequence, the average velocity of the ratchet reads~\cite{ref21}
\begin{multline}
\langle \omega \rangle = \frac{\rho S}{2 \dat}\int dv |v| \phi(v)\Theta(-v) \int
\frac{ds}{S} \sigma(g) \frac{v^2}{\ri^2}\frac{(1+\alpha)^2\eps^4 g^2}{(1+\eps^2
g^2)^2}= \\\frac{\rho S}{\sqrt{2 \pi}\dat}\frac{\eps^4 v_0^3}{\ri^2}
\left\langle
(1+\alpha)^2 \sigma(g)\frac{g^2}{(1+\eps^2 g^2)^2} \right\rangle_{surf}.
\end{multline}
where we have used the shorthand notation for the average along the perimeter of
the rotator's shape $\langle \rangle_{surf} = \int \frac{ds}{S}$.
If one considers the expression for the control parameter $\beta^{-1}$
 we can finally write for the adimensional
angular velocity $\Omega = \frac{\ri}{\eps v_0}\omega$
\begin{equation} 
\langle \Omega \rangle  =  \sqrt{\pi}(1+\alpha)^2\beta^{-1} \eps^2 \left\langle
\frac{ \sigma(g)g^2}{(1+\eps^2 g^2)^2} \right\rangle_{surf}.
\end{equation}

\subsection{Details of the simulations}

We have simulated the model described in equations~(2) of the Letter, through
a suitable adaptation of the Direct Simulation Monte Carlo method
(DSMC)~\cite{ref31}. The DSMC is devised to solve numerically a Boltzmann
equation, therefore enforcing the Molecular Chaos assumption. In our
specific problem, however, the procedure is drastically simplified,
since only one particle (the rotator) is represented in the
simulation, through its angular velocity and position, $\omega$ and
$\theta$ respectively. The surrounding gas is represented by its
constant velocity distribution $\phi({\bf v})$ (assumed Gaussian with
variance $v_0^2$), unaffected by collisions.  The dynamics of these
variables advances by a series of constant and small time steps of
length $\delta t$ taken smaller than all characteristic time-scales in
the problem. At every time step a {\em free streaming} update and a
{\em collisional} update are performed. The free streaming corresponds
to the evolution of $\omega$ and $\theta$ from $t$ to $t+\delta t$ in
the absence of any interrupting collisions. In the {\em collisional}
update, a collision is performed with the probability dictated by the
Boltzmann equation (Eq. (3) of the Letter). The correct probability is sampled
through a Monte Carlo procedure, where a {\em tentative} collision is
proposed, by choosing the point of impact (at random and uniformly)
along the surface of the rotator and by extracting the velocity ${\bf
  v}$ with probability $\phi({\bf v})$. The {\em tentative} collision
is realized with a probability equal to $p_c=(\bV- \bv) \cdot \hatn
\rho S \delta t$ (note that $\delta t$ is chosen much smaller than
$\tau_c$ and this guarantees that $p_c \ll 1$), leading to an update
of $\omega$ by the collision rule. Otherwise no collisions occur. Note that the choice of
the point of impact (which determines also $\hatn$) can reproduce the
precise shape of the experimental rotator (symmetric or chiral).

{\em Control parameter $\beta^{-1}$ as a good estimate of the ratio of typical times.}
In Figure~5 we report the ratio between the collision
times $\tau_c$ and frictional stop times $\tau_\Delta$, observed in
the simulations in different regimes, in order to assess the fairness
of estimate $\beta^{-1} \approx \tau_\Delta/\tau_c$.

\begin{figure}[h]
\includegraphics[width=8cm,clip=true]{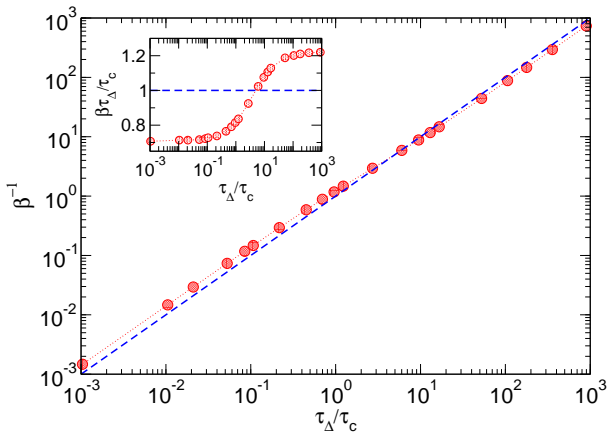}
\caption{ Assessment of the estimate $\beta^{-1}$ for
  $\tau_\Delta/\tau_c$. The ratio between $\tau_\Delta/\tau_c$ and
  $\beta^{-1}$ obtained in numerical simulations is shown in the inset. The estimate is useful in
  experiments, where measurements of $\tau_\Delta$ and $\tau_c$ are
  not reliable. It is seen that the estimate is a good approximation
  of the time-scale ratio where it is close to $1$, while it slightly
  deviates from it in the RCL and FCL, by a factor $\sim 1.25$ and
  $\sim 0.7$ respectively.}
\end{figure}

\subsection{Experimental set-up.}

We assume that the evolution of the angular velocity $\omega$
of the rotator obeys an equation of the form
\begin{equation}
\label{motioneq}
I\dot{\omega}(t)=-F_{friction}\sigma[\omega(t)]-\Gamma_{visc}\omega(t)+F_{coll}
(t),
\end{equation} where $I$ is the momentum of inertia, 
$F_{friction}$ is the dry friction torque, $\sigma$ the sign
function and $\Gamma_{visc}$ the air viscous drag. The term $F_{coll}=I\eta_{coll}$
contains the driving torque, that is the angular momentum randomly transferred
from the beads to the rotator, and is analysed in detail in
Section 2.  We refer to Figure 3a of the paper for the main
features of our set-up. Dissipation coefficients rescaled by inertia
are $\gamma_a=\Gamma_{visc}/I$ and $\Delta=F_{friction}/I$.

{\em Technical details about the experimental setup}. The granular
medium, made of $50$ polyoxymethylene (POM) spheres (radius $r=3 $~mm
and mass $m=0.15$~g), is housed in a polymethyl-methacrylate (PMMA)
cylinder (diameter $90$~mm) with a conical-shaped floor.  A removable
cap encloses a miniaturized angular encoder (model AEDA-3300 by Avago
Technologies). The encoder, which also supports the rotator, provides
high resolution measurements (up to 80,000 division/revolution at the
maximum rate of $20$~kHz) of the rotator position.  The system is
vibrated by an electrodynamic shaker (model V450 by LDS Test \&
Measurement) fed by a sinusoidal excitation. An accelerometer measures
the actual acceleration induced to the system.  A high-speed camera
(EoSens CL by Mikrotron) tracks single beads in order to measure their
velocity.  The two rotators are made of PMMA, have height $h=15$~mm
and are distinguished by their different section on the rotation
plane. The momentum of inertia of the rotator $I$ comprises different
parts.  The probe is attached to the angular encoder by means of a
steel rotation axis $50$~mm high and $3$~mm thick.  The mass of the whole
rotator (axis and probe) is $M=5.21$~g for the asymmetric probe and
$M=6.49$~g for the symmetric one.  The total momentum of inertia is
$I=135$~g\,mm$^2$ and $I=353$~g\,mm$^2$ for the two types,
respectively.  In both cases the inertia of the axis is a few
hundredths of the total one. The shape factors of the asymmetric
rotator, relevant for its ratchet effect, are $\left\langle
\frac{\sigma(g)g^2}{(1+\eps^2g^2)}\right\rangle_{surf}=0.0013$ and
$\frac{\langle g^3 \rangle_{surf}}{\langle g^2 \rangle_{surf}}=0.052$
for the RCL and FCL limits respectively (see text for
definitions). Both the shape factors vanish for the symmetric rotator.

{\em Dry and viscous friction}. The only source of dry friction comes
from the two ball bearings inside the encoder while viscous friction
is due to the air surrounding the rotator.  We measure the dry and
viscous friction during a standard experimental run (eight hours), but
considering only periods where the rotator is subject to these forces
and no other: $\dot{\om}=-{\gat}\om \mp \dat$ (where the upper sign
holds for $\om>0$ and the lower for $\om<0$), i.e. excluding the time
immediately about collisions and periods of rest. Angular velocity and
acceleration are obtained from the angular position (in function of
time) by differentiation. A bidimensional histogram is calculated and
a peak-finding procedure has been applied in order to obtain the most
probable trajectories in ($\om, \dot{\om}$) space. Finally, the most
probable trajectory is fitted with the above linear equations,
yielding the values $F_{friction}=(9.9 \pm 0.7) \times 10^3
$~g\,mm$^2$/s$^2$ and $\Gamma_{visc}=(1.6\pm 0.2) \times 10^3$~g\,mm$^2$/s.  It is interesting to notice that viscous friction
becomes important for velocities larger than the threshold velocity
$\omega_{th}=F_{friction}/\Gamma_{visc} \approx 6.2$~s$^{-1}$,
separating the friction-dominated regime $\omega \ll \omega_{th}$ from
the viscosity-dominated one $\omega \gg \omega_{th}$.

{\em Restitution coefficients}.
The beads-rotator restitution coefficient $\alpha$ has been measured
launching a single bead against the rotator while recording the
rotator position at high sampling rate (1 kHz). For these measurements
the top of the apparatus is removed and put with the rotation axis
parallel to the floor. A bead falls from height $h$ and hits the
rotator with velocity $v=\sqrt{2 a_g h}$ (here $a_g$ indicates the
gravity acceleration). The high-speed camera has been used to monitor
the exact distance $x$ of the impact point from the rotation axis. We
calculate the restitution coefficient using the collision rule (see
Section 2 below) adapted for this particular configuration in which
the rotator is at rest before the collision ($\Om=V=0$) and $g\approx
x/R_{I}$. In this circumstances the velocity of the rotator after a collision is
$\om'=(1+\alpha)vx/(I/m+x^2)$. Repeated measurements on the symmetric
rotator gave the following results: $\alpha=0.83 \pm 0.16$ (we recall
that $\alpha=1$ for elastic collisions).

{\em Granular gas velocity statistics}.
We discuss here the methods used to characterize the velocity
statistics of the granular gas. A fast camera (EoSens CL by Mikrotron)
is placed above the PMMA container to catch horizontal components of
motion $(v_x,v_y)$ of the polyoxymethylene (POM) spheres constituting
the granular gas.  Focus of lens is adjusted to the plane at half
height of the probe. Lighting is provided by a led-dome which diffuses
light on the system reducing shadows and reflections.  Particle tracking
is enhanced by marking a few tracers: $3$ POM spheres are white, while
all the other particles are black. Pictures are taken at $250$ frames
per second: this is verified to be the optimal compromise between too
large delays, which prevent tracking of ballistic trajectories, and
too small ones, which produce ``false movement'' induced by noise due
to finite sensitivity of the camera resolution and the centring
algorithm. Uncertainty in the determination of the centre of mass of
tracer spheres is estimated to be $\sim 0.05$ mm.  The details of
velocity measurements and error estimates are reported in~\cite{gnoli}. In
Figure~6 we display a typical snapshot of the system
from the camera (see also the accompanying Movie).  

\begin{figure}[h]
\includegraphics[width=7cm,clip=true]{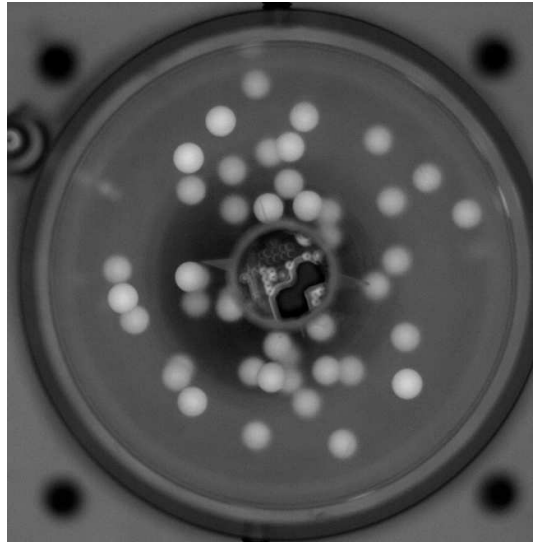}
\caption{A snapshot of the system taken from the fast camera (shaker
  is vibrating at 53 Hz with rescaled maximum acceleration
  $\Gamma=8$). }
\end{figure}

For each choice of shaking parameters we have computed the velocity
histogram from a series of $2\times 10^4$ contiguous frames. Examples
of those histograms are reported in Figure~7, left frame. A Gaussian
fit is seen to be a good approximation and provides the value of $v_0
= \sqrt{\langle v^2 \rangle}$ where $v$ is $v_x$ or $v_y$ (isotropy is
always verified). In the right frame of Figure~7 we report the values
of $v_0$ as function of the rescaled maximum acceleration
$\Gamma=\frac{a_{max}}{a_g}$ where $a_g$ is gravity acceleration and
the $z_s(t)$ position of the shaker's vibrating head follows the law
$z_s(t)=\frac{a_{max}}{(2\pi f)^2}\sin(2\pi f t)$ with $f$ being the
vibration frequency.

\begin{figure}[h]
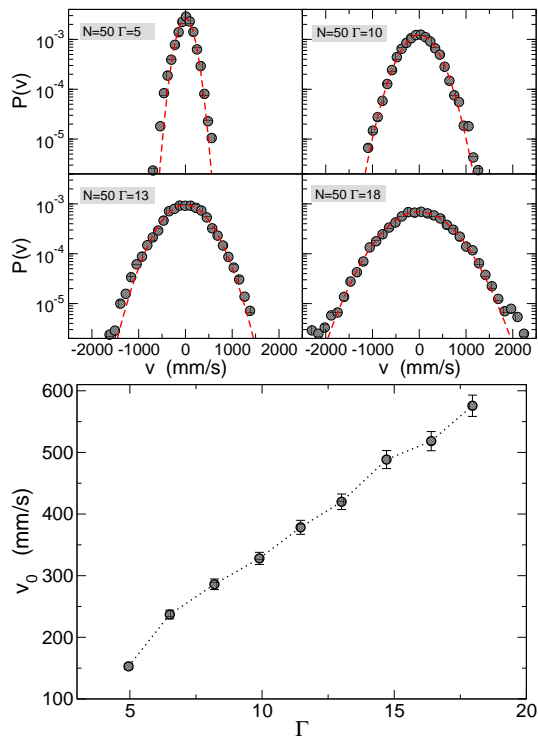

\includegraphics[width=7cm,clip=true]{vpdf.eps}
\includegraphics[width=7cm,clip=true]{v0.eps}
\caption{ Left: Histograms of velocities ($v_x$) of $N=50$ granular gas
  particles for different values of the rescaled maximum acceleration $\Gamma$
  at vibration frequency 53~Hz. Gaussian fits are
  superimposed. Right: the values of $v_0$ (standard deviation) versus
  $\Gamma$. }
\end{figure}

{\em Rotator velocity signal}.
An example of the velocity signal $\omega(t)$ in two different regimes
(frequent and rare collisions) is shown in Figure~8, left frame, for
the case of the symmetric rotator.  In the right frame of the same
Figure, we show the angular velocity autocorrelation $C(t)=\langle
\omega(t) \omega(0) \rangle - \langle \omega \rangle^2$ versus
time. Note that the signal average is much smaller than its standard
deviations and, hence, it is negligible in $C(t)$.

\begin{figure}
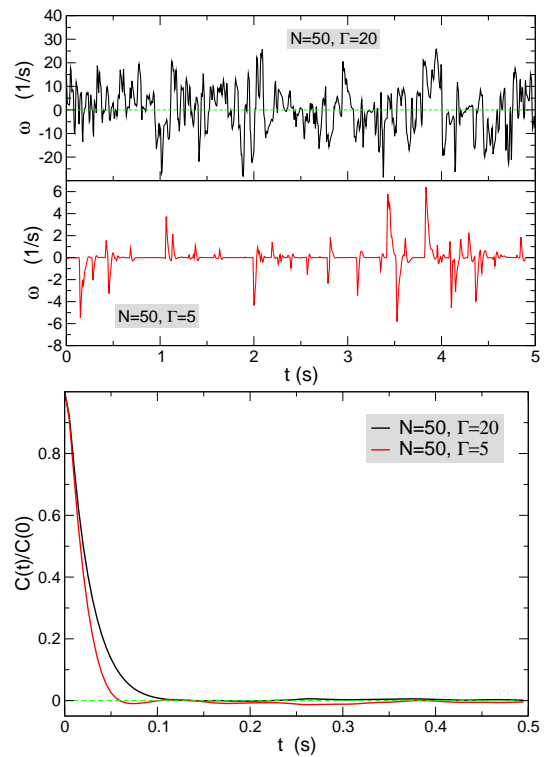

\includegraphics[width=7cm,clip=true]{signal.eps}
\includegraphics[width=7cm,clip=true]{c.eps}
\caption{ Left: Examples of symmetrical rotator's angular velocity signal
$\omega(t)$  as acquired from the angular encoder at 200 Hz in two different
  regimes (frequent collisions, top, and rare
  collisions, bottom).  Right: rescaled angular velocity
  autocorrelation $C(t)/C(0)$ versus time $t$ for the two cases shown
  in the left frame.}
\end{figure}

Here, it is interesting to comment about the smaller characteristic
time associated to the system with rare collisions (red curve), with
respect to that with frequent collisions (black curve): this is a
consequence of the smaller typical velocity in the rare collisions
configuration, which induces faster relaxations through Coulomb
friction dissipation (we recall that $\tau_\Delta \approx
\frac{\langle|\omega|\rangle}{\Delta}$).  Similar results are obtained
with the asymmetric rotator.


\end{document}